\documentclass[11pt,letterpaper,oneside]{article}
\usepackage[letterpaper,margin=1in]{geometry}

\usepackage{setspace}
\usepackage{amsmath,amssymb,amsthm,mathtools}
\usepackage[
      backend=biber,
      style=numeric,
      sorting=none,
      maxbibnames=99,
      doi=false,
      url=false,
      isbn=false,
      eprint=true
  ]{biblatex}
  \AtEveryBibitem{%
    \clearname{editor}%
    \clearfield{eventtitle}%
    \clearfield{eventtitleaddon}%
    \clearfield{venue}%
    \clearfield{eventdate}%
    \clearfield{booktitleaddon}%
    \clearlist{organization}%
    \clearlist{publisher}%
  }
\usepackage{algorithm,algpseudocode}
\usepackage{tabularx,booktabs,array}
\usepackage{enumitem}
\usepackage{xcolor}
\usepackage{tikz}
\usetikzlibrary{arrows.meta, patterns}
\usepackage{microtype}
\usepackage{url}
\usepackage[hidelinks]{hyperref}
\usepackage[nameinlink,noabbrev]{cleveref}
\usepackage{tikz}
\usetikzlibrary{arrows.meta,positioning,shapes.geometric}
\makeatletter
\providecommand*{\theHALG@line}{}
\makeatother

\newcommand{\argmin}{\operatorname*{arg\,min}}
\newcommand{\rat}{\operatorname{rat}}

\newcommand{\MFCP}{\textnormal{MFCP}}
\newcommand{\SSR}{\textnormal{SSR}}
\newcommand{\APR}{\textnormal{APR}}
\newcommand{\BRBI}{\textnormal{BRBI}}
\newtheorem{example}{Example}
\usepackage{setspace}
\algrenewcommand\algorithmiccomment[1]{\hfill\(\triangleright\) #1}
\algrenewcommand\algorithmicindent{0.9em}
\setlist{nosep,leftmargin=*}
\allowdisplaybreaks[2]
\hypersetup{pdftitle={Polynomial-Time Approximate EFX for Indivisible Chores},pdfauthor={Author information to be completed}}

\floatname{algorithm}{Algorithm}
\algrenewcommand\algorithmicrequire{\textbf{Input:}}
\algrenewcommand\algorithmicensure{\textbf{Output:}}
\algrenewcommand\algorithmicif{\textbf{if}}
\algrenewcommand\algorithmicthen{\textbf{then}}
\algrenewcommand\algorithmicelse{\textbf{else}}
\algrenewcommand\algorithmicfor{\textbf{for}}
\algrenewcommand\algorithmicdo{\textbf{do}}
\algrenewcommand\algorithmicend{\textbf{end}}
\algrenewcommand\algorithmicreturn{\textbf{return}}
\algrenewcommand\algorithmicloop{\textbf{loop}}
\theoremstyle{plain}
\newtheorem{theorem}{Theorem}[section]
\newtheorem{lemma}[theorem]{Lemma}
\newtheorem{corollary}[theorem]{Corollary}
\newtheorem{proposition}[theorem]{Proposition}

\newtheorem{definition}[theorem]{Definition}

\tikzset{
phasebox/.style={draw, rounded corners=2pt, align=center, minimum width=2.5cm, minimum height=0.62cm, inner sep=2pt},
phaseline/.style={-{Latex[length=2mm]}, thick},
smalllabel/.style={font=\scriptsize, align=center}
}

\begin{document}

\begin{center}
    {\LARGE\bfseries
    Polynomial-Time Approximate EFX for Chores via Flow and Matching
    \par}

    \vspace{0.5em}

    {\large
    Genjie Qin$^{1}$, Qizhi Fang$^{1}$,
    Minming Li$^{1,2}$,
    Wenjing Liu$^{1,3}$, Xin Chen$^{4,*}$
    \par}

    \vspace{0.35em}

    {\normalsize
    $^{1}$School of Mathematical Sciences, Ocean University of China\\
    $^{2}$Department of Computer Science, City University of Hong Kong\\
    $^{3}$School of Mathematical Sciences and Laboratory of Marine Mathematics,\\
    Ocean University of China\\
    $^{4}$College of Computer Science and Technology, Ocean University of China
    \par}

    \vspace{0.3em}

    {\normalsize
    Genjie Qin: \texttt{qgj@ouc.edu.cn};
    Qizhi Fang: \texttt{qfang@ouc.edu.cn}\\
    Minming Li: \texttt{minming.li@cityu.edu.hk};
    Wenjing Liu: \texttt{liuwj@ouc.edu.cn}\\
    Xin Chen$^{*}$: \texttt{chenxin1403@ouc.edu.cn}
    \par}

    \vspace{0.2em}
    {\normalsize $^{*}$ Corresponding author.}
\end{center}

\vspace{0.3cm}

\begingroup
\setlength{\parindent}{0pt}  
\setlength{\parskip}{3pt}    

\noindent\textbf{Abstract}
We study approximate EFX allocations of indivisible chores under nonnegative additive costs. Exact EFX allocations need not exist in general. We develop a polynomial-time algorithm that computes a complete $(3+\varepsilon)$-EFX allocation for any rational $0<\varepsilon\le1$, improving the previous approximation guarantee from a factor depending on the number of agents to a constant. The algorithm extends to EF$k$X, achieving a $(2+1/k+\varepsilon)$-approximation for every integer $k\ge1$.

For instances with relatively few chores, the polynomial-time BRBI algorithm gives stronger guarantees. It finds exact EF$k$X allocations when $m\le2kn$ and achieves a factor of at most $2-1/k$ when $m\le(3k-1)n$. In particular, $k=1$ and $m\le2n$ give exact EFX.

We further show that the approximation factor \(2+1/k\) is asymptotically tight for our main algorithm. We also construct a four-agent, ten-chore instance in which every complete allocation has EFX approximation factor at least $11/10$, improving upon the $20/19$ lower bound for submodular costs even under additive costs.


\endgroup

\newpage
\section{Introduction}\label{sec:intro}

Fair allocation of indivisible chores concerns the distribution of burdens among agents with heterogeneous preferences and is a fundamental problem in algorithmic game theory and social choice~\cite{AzizSurvey2022}.
We consider a set $N$ of $n$ agents and a set $M$ of $m$ indivisible chores, which impose costs on their recipients.
Each agent $i$ has a nonnegative additive cost function $c_i(\cdot)$, with $c_i(S)=\sum_{g\in S}c_i(g)$ for every bundle $S\subseteq M$.
A complete allocation $X=(X_i)_{i\in N}$ assigns every chore to exactly one agent, with $X_i$ denoting the possibly empty bundle assigned to agent $i$.
Envy-freeness (EF) would require each agent to
weakly prefer her own bundle to every other bundle, but even one positive-cost chore can make this impossible. 
Envy-freeness up to any chore (EFX) asks that envy disappear after any one chore is removed from the agent's own bundle, with both bundles evaluated using that agent's costs.
Our goal is to compute, in polynomial time, allocations satisfying approximate EFX.
For $\alpha\geq 1$, an allocation is $\alpha$-EFX if, after removing any single chore from an agent's own bundle, her remaining cost is at most $\alpha$ times her cost for any other agent's bundle.

Exact EFX allocations can be computed in polynomial time in several restricted settings, including $m\leq 2n$~\cite{KMS2025}. 
In general, however, exact EFX allocations need not exist, even when all costs are drawn from a set of three positive values~\cite{HT2026}.
We study multiplicative approximation: for $\alpha\geq 1$, an allocation is $\alpha$-EFX if removing any single chore from $X_i$ leaves agent $i$ with cost at most $\alpha\cdot c_i(X_j)$ for every other agent $j$.
Garg et al.~\cite{GMQ2025} established the existence of $4$-EFX allocations using earning-restricted equilibria, and Garg and Murhekar~\cite{GM2026} subsequently improved the guarantee to $2$ via local swaps on a suitable price-supported EF1 allocation.
On the algorithmic side, Zhou and Wu~\cite{ZhouWu2024} presented a polynomial-time $(3n^2-n)$-EFX algorithm for $n\ge 4$, but its approximation factor grows quadratically with $n$. We therefore ask whether a constant-factor EFX allocation can be computed directly from the cost matrix in polynomial time.

\subsection{Results}\label{sec:contributions}

We present a polynomial-time $(3+\varepsilon)$-EFX algorithm for general additive chores in Section~\ref{sec:main-algorithm}, reducing the quadratic dependence on $n$ in the guarantee of Zhou and Wu~\cite{ZhouWu2024} to a constant. 
Let $L_c$ denote the total binary encoding length of the input cost matrix.
\begin{theorem}\label{thm:main}
For every rational $0<\varepsilon\le1$, a complete $(3+\varepsilon)$-EFX allocation can be computed in time polynomial in $n$, $m$, $L_c$, and $1/\varepsilon$.
\end{theorem}
To the best of our knowledge, this is the first polynomial-time constant-factor EFX algorithm for general additive chores. The approximation factor is independent of the cost magnitudes and ratios. Although a stronger $2$-EFX existence guarantee is known~\cite{GM2026}, Theorem~\ref{thm:main} makes a constant-factor approximation efficiently computable.

A key algorithmic contribution of our work is Max-Flow Capped Price (MFCP), which constructs a capped-price structure directly from the input cost matrix. By combining max-flow feasibility with min-cut-guided multiplicative weight updates, MFCP obtains a complete fractional allocation with balanced capped incomes and approximate MPB support in polynomial time.

Our main focus is EFX, but the construction also extends to EF$k$X. For an integer $k \geq 1$, EF$k$X requires each agent $i$ to be envy-free after any $\min\{k, |X_i|\}$ chores are removed from her own bundle.
This differs from EF$k$, which asks only for some deletion of at most $k$ chores in each comparison. 
Under additive costs, it suffices to remove the $\min\{k, |X_i|\}$ cheapest chores according to $i$, since this leaves the largest remaining cost.

\begin{corollary}\label{cor:main-generalization}
For every integer $k\geq 1$ and rational $0<\varepsilon\le1$, a complete $(2+1/k+\varepsilon)$-EF$k$X allocation can be computed in time polynomial in $n$, $m$, $L_c$, and $1/\varepsilon$.
\end{corollary}

When there are relatively few chores, a simpler balanced reservation and block insertion (BRBI) yields a stronger guarantee. 
In Section~\ref{sec:brbi} and Appendix~\ref{app:brbi-proof}, we show that BRBI also runs in polynomial time and computes a $\max\!\left\{1,\frac{\lceil(m-kn)/n\rceil}{k}\right\}$-EF$k$X allocation.
This yields an exact EF$k$X allocation when $m\le2kn$, and an approximation factor of at most $2-1/k$ when $m\le(3k-1)n$.
While the exact EFX guarantee for $m\leq 2n$ is already known~\cite{GM2026,GMQ2025,KMS2025}, 
we extend this approach to base bundles of $k$ chores and analyze the insertion of the remaining chores.

Furthermore, we present an algorithmic tightness example and an EFX existence lower bound.
For the former, we construct, for every $k\ge1$, a family of instances on which the EF$k$X approximation ratio of our main algorithm's output approaches $2+1/k$.
Thus the leading factor is asymptotically tight for our algorithm, reaching $3$ for EFX.
\begin{theorem}\label{thm:efx-lower-bound-main}
There exists an instance with four agents, ten indivisible chores, and nonnegative additive costs in which every complete allocation has EFX approximation factor at least $11/10$.
\end{theorem}

\subsection{Algorithmic Overview}\label{sec:story}

Our main algorithm (Algorithm~\ref{alg:main}) handles the single-agent case directly and invokes BRBI when $m \leq 2n$. For the remaining instances, it proceeds according to the following framework:
\[
 (\widehat C_i(g))_{i\in N,g\in M}
 \xrightarrow{\ \text{\normalsize MFCP}\ } \mbox{fractional~} (w_{ig})_{i\in N,g\in M}
 \xrightarrow{\ \text{\normalsize SSR}\ }  Y
 \xrightarrow{\ \text{\normalsize APR}\ }\mbox{integral~}X.
\]
Before these stages, we construct a strictly positive
working-cost matrix
$\widehat C=(\widehat C_i(g))_{i\in N,g\in M}$
from the input cost matrix
$c=(c_i(g))_{i\in N,g\in M}$.
MFCP, SSR, and APR all operate on $\widehat C$,
and the approximation guarantee holds for both \(\widehat C\) and the original costs \(c\).
\begin{figure}[htbp]
  \centering
  \includegraphics[width=1\linewidth]{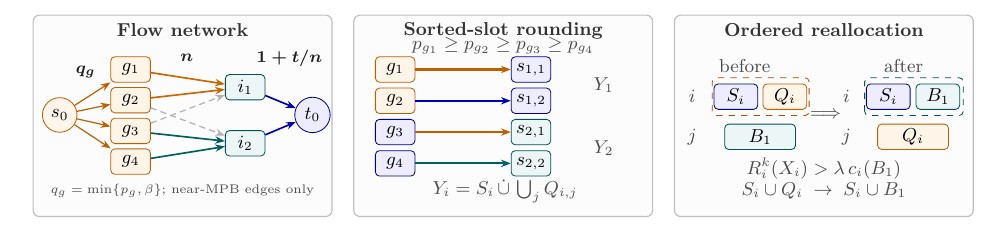}
  \par\vspace{.25em}
  \begin{minipage}[t]{.317\textwidth}
  \raggedright \footnotesize
  \textbf{(a) MFCP.} Flow network with capped capacities and eligible edges.
  \end{minipage}\hfill
  \begin{minipage}[t]{.317\textwidth}
  \raggedright \footnotesize
  \textbf{(b) SSR.} Price-ordered matching of chores to agent slots.
  \end{minipage}\hfill
  \begin{minipage}[t]{.317\textwidth}
  \raggedright \footnotesize
  \textbf{(c) APR.} Ordered packet reallocation for envy repair.
  \end{minipage}
  \caption{Overview of the three stages of the EFX algorithm.}
  \label{fig:three-phases}
\end{figure}
\paragraph{Max-Flow Capped Price} 
The first challenge is to compute a useful price support in polynomial time.
Inspired by the earning-restricted approach~\cite{GMQ2025}, we assign positive chore prices $p_g$ and cap their income contributions at $q_g=\min\{p_g,\beta\}$.
Algorithm \textsc{MFCP} updates agent weights $v_i$ and scales the base prices $r_g(v)=\min_{i\in N}v_i\widehat C_i(g)$ such that the total capped income equals $n$.
We construct a flow network and compute a maximum flow. If its value is $n$, the flow yields a complete fractional allocation supported on approximately MPB edges.
Otherwise, a minimum cut identifies the agents whose weights are to be increased by a factor of $(1+t)$, and the prices are recomputed accordingly. This iterative procedure terminates in polynomial time and yields a fractional allocation satisfying the required conditions.

\paragraph{Rounding}
Algorithm \textsc{SSR} extends the Shmoys--Tardos rounding framework~\cite{ST1993} by assigning fractional shares to unit slots in nonincreasing order of capped prices, placing chores with \(p_g>\beta\) first among ties.
Since each agent's fractional mass on these chores is at most \((1+t)/\beta<2\), they occupy at most two slots. After padding partial slots with zero-price dummy chores, we find a perfect matching in the support graph, which yields an integral allocation \(Y\).
The matching preserves approximate MPB support, loses at most \(\beta\) in capped income, and assigns at most two chores with \(p_g>\beta\) to each agent.
For each agent \(i\), we separate the chores with \(p_g>\beta\) in \(Y_i\) as singleton packets and let the remaining chores form \(S_i\). Let \(\nu_i\le2\) denote the number of packets.
Each packet has price greater than \(\beta\) and zero EFX residual when assigned alone. Since \(p_g=q_g\) for all \(g\in S_i\), we have \(p(S_i)=q(Y_i)-\nu_i\beta\). Ignoring approximation slack, each \(Y_i\) and each packet has price at least \(1/2\), while \(p(S_i)\le(3-\nu_i)/2\).

\paragraph{Envy Repair}
Finally, Algorithm \textsc{APR} repairs the remaining fairness violations without undoing earlier repairs.
Active agents choose their cheapest available packets in a fixed order, and a second pass checks and repairs their allocations. 
The order ensures that packets chosen later are not cheaper to an earlier agent than the packets she could choose.
A recipient of a pure packet receives at most $k$ chores and therefore has zero residual cost after deleting $k$ chores; a composite bundle contains an earlier complete bundle and so cannot become cheaper to other observers. These properties prevent later exchanges from undoing earlier repairs.
These stages together establish the required fairness conditions. The flow stage provides cost-to-price comparability, matching controls rounding loss and the number of high-price chores, and the final exchange converts these bounds into fairness against every comparison bundle.

\subsection{Related work}\label{sec:related}

Exact EFX is known for several restricted chore settings, including two-type and binary costs, and for a bounded number of agents or chores \cite{AzizEtAl2023,TaoWuYuZhou2025,Zhang2026}; a $2$-EFX algorithm is also known for three agents \cite{Afshinmehr2024}.
For general additive chores, Zhou and Wu~\cite{ZhouWu2024} gave a polynomial-time $(3n^2-n)$-EFX algorithm for $n\ge4$ (with running time $O(nm\log m)$). Garg, Murhekar, and Qin~\cite{GMQ2025} subsequently established the existence of $4$-EFX allocations via earning-restricted (ER) equilibria and rounding. Their reallocation procedure runs in polynomial time given an ER equilibrium; their general approach to computing that equilibrium enumerates support graphs in $m^{O(n^2)}$ time, which is polynomial only for fixed $n$.
This does not establish an exponential-time lower bound for computing $4$-EFX. Garg and Murhekar~\cite{GM2026} later improved the general existence factor to $2$-EFX, again without a polynomial-time procedure for obtaining the required starting allocation in general.
Christoforidis~\cite{Christoforidis2026} showed that, under monotone submodular costs, there exists a three-agent, six-chore instance admitting no $\alpha$-EFX allocation for any $1\leq\alpha<20/19$. 
Related work studies bivalued costs, EF1, fractional Pareto optimality, and approximate competitive equilibria \cite{LinWuZhou2025,Bhaskar2020,GMQ2022,Mahara2025,Boodaghians2022}. Our algorithm constructs its approximate price certificate directly from the input rather than assuming an exact equilibrium or a specially structured integral starting allocation.

EF$k$X has also attracted interest. 
Originally introduced for goods by Akrami et al.~\cite{ARS2022} who achieved exact EF$2$X for restricted additive valuations. 
Ashuri et al.~\cite{AGS2025} obtained exact EF$2$X for four agents with cancelable valuations in pseudopolynomial time.
EF$k$X has also been studied for chores.
For binary supermodular costs, Barman et al.~\cite{BNV2023} showed that there are instances in which no Pareto-optimal allocation achieves any finite EF$k$X approximation factor.
Our work instead focuses on additive costs and provides fairness guarantees.


\section{Preliminaries}\label{sec:model}

Consider $m$ indivisible chores (forming a set $M$) and $n\geq 2$ agents (forming a set $N$). 
For each $i\in N$ and each $g \in M$, denote $c_i(g)$ the cost of agent $i$ on chore $g$. We assume that $c_i: 2^M \rightarrow \mathbb{Q}_{\geq 0}$ is an additive function with $c_i(\emptyset) = 0$.
For our algorithmic results, the input is a cost matrix $(c_i(g))_{i\in N,g\in M}$.
A complete (integral) allocation $X = (X_i)_{i\in N}$ is a partition of $M$ into $n$ possibly empty bundles. 
Our main goal is to compute an allocation satisfying $\alpha$-EFX in polynomial time.

\begin{definition}[$\alpha$-EFX]\label{def:fairness}
A complete allocation $X$ is $\alpha$-EFX if, for every $i,j\in N$ and every $g\in X_i$,
$c_i(X_i\setminus\{g\})\leq \alpha \cdot c_i(X_j)$.
When $\alpha = 1$, we call the allocation exact EFX. 
\end{definition}
We also consider the extension $\alpha$-EF$k$X.
For any $k\geq 1$ chores, an allocation $X$ is $\alpha$-EF$k$X if $R_i^k(X_i)\leq \alpha\cdot c_i(X_j)$ for all $i$ and $j$, where $R^k_i(S) = \max\{c_i(S\setminus T): T\subseteq S, |T| = \min\{k, |S|\}\}$.
In particular, $R_i^k(S)=0$ whenever $|S|\leq k$, and $\alpha$-EF$1$X coincides with $\alpha$-EFX. For a complete allocation $X$, we define its EF$k$X approximation ratio as
\(
\rat_k(X)=\inf\{\alpha\ge1:
X\text{ is }\alpha\text{-EF}k\text{X}\}\) and \(
\rat(X)=\rat_1(X).
\)

\paragraph{Prices and MPB}
Recall that the first stage of our main algorithm is to compute a fractional solution $w = (w_{ig})_{i\in N, g\in M}$ satisfying that every $w_{ig} \geq 0$ and $\sum_{i\in N} w_{ig} = 1$ for every $g$.

To construct the fractional $w$, we associate each chore $g\in M$ with a price $p_g >0$.
Following the Fisher-market interpretation~\cite{GMQ2025}, $p_g$ can be thought of as the payment for performing one unit of chore $g$.
For a positive additive cost matrix $(c_i(g))_{i\in N, g\in M}$, associate each chore
$g\in M$ with a positive price $p_g$. Define agent $i$'s minimum pain per buck (MPB)
ratio as
\(
\gamma_i=\min\{c_i(g)/p_g: g\in M\}.
\)
For $a\ge1$, a fractional allocation $w$ is
$a$-MPB-supported if $w_{ig}>0$ implies
$g\in A_i(a)$, where
\(
A_i(a)=\{g\in M:c_i(g)\le a\gamma_i p_g\}.
\)
For an integral allocation $Y$, this reduces to $Y_i\subseteq A_i(a)$.
\begin{proposition}\label{prop:mpb}
Fix a positive price vector $p$ and an agent $i$.  For every $B\subseteq M$ and every $S\subseteq A_i(a)$, the following hold. 
\begin{align*}
(1) &~~ c_i(B) \geq \gamma_i \cdot p(B),~~ c_i(S) \leq a\cdot \gamma_i \cdot p(S);\\
(2) &~~ \text{If~} p(B) \geq \rho > 0, ~p(S) \leq b, \text{then~} c_i(S) \leq ab/\rho \cdot c_i(B). ~~~~~~~~~~~~~~~~~~~~~~~~~~~~~~~~~~~~~~~~~~~~~
\end{align*}
\end{proposition}

Subsequently, Algorithm~\ref{alg:flow} constructs a fractional allocation with support factor $a = 1 + t$. Algorithm SSR then performs rounding within this support, so the residual $S_i \subseteq Y_i$ left after packet extraction inherits the cost upper bound. 
Thus, Proposition~\ref{prop:mpb} allows us to bound an agent's cost for an $a$-MPB-supported bundle $S$ relative to any comparison bundle $B$, using only an upper bound $b$ on $p(S)$ and a positive lower bound $\rho$ on $p(B)$.

\paragraph{Capped Price} 
To balance fractional income and control rounding loads, we introduce capped price vector $q$. Specifically, for a cap $\beta >0$, let $q_g = \min\{ p_g, \beta \}$ and $q(B) = \sum_{g\in B} q_g$. For each agent $i$, her total capped income is defined as $E_i = \sum_{g\in M} q_g\cdot w_{ig}$.
For a strictly positive additive cost matrix $c$, we define the following capped-price structure.
\begin{definition}[Capped-price structure]\label{def:capped-price}
For $0< t \leq 1/100$ and $0<\beta < 1$, a capped-price structure consists of positive prices $p_g$, MPB ratios $\gamma_i$ and fractional shares $w_{ig}$ satisfying
\begin{align}
(1) &~~ \sum_{i\in N} w_{ig} = 1, ~~\forall g\in M; \tag{completeness}\\
(2) &~~  E_i \in[1 - t, 1 +t],  ~~\forall i\in N;  \tag{capped-income balance}\\
(3) &~~ c_i(g) \geq \gamma_i \cdot p_g, ~~\forall i\in N, g\in M; \tag{cost lower bound} \\
(4) &~~  w_{ig}>0 \Longrightarrow\ c_i(g)\le(1+t)\gamma_i\cdot p_g.~~~~~~~~~~~~~~~~~~~~~~~~~~~~~~~~~~ \tag{approximate MPB support}
\end{align}
\end{definition}
In the main algorithm, we apply this structure to the positive working-cost matrix $\widehat C$. Notice that this structure allows small errors in support and income balance, while its complete representation permits every chore to enter rounding. 
It relates to earning-restricted (ER) equilibrium~\cite{GMQ2025} through $x_{ig} = (q_g/p_g) \cdot w_{ig}$ satisfying $\sum_{i\in N} x_{ig}= \min\{1, \beta/ p_g\}$ and $\sum_{g\in M} p_g \cdot x_{ig} = E_i$. 
It is also related to competitive equilibrium with equal incomes (CEEI), which combines exact MPB support, complete fractional allocation, and uncapped income $\sum_{g} p_g\cdot x_{ig} = 1$.
Although approximate CEEI admits an FPTAS~\cite{Boodaghians2022}, its guarantee does not directly provide our capped structure.

\section{Main Algorithm}\label{sec:main-algorithm}

Our overall algorithm (Algorithm~\ref{alg:main}) first distinguishes sparse instances from general instances.
If $m \leq 2n$, the BRBI procedure (in Section~\ref{sec:brbi}) directly constructs an exact EFX allocation. 
For instances with $m > 2n$, define the positive working costs
$\widehat C_i$ as follows. If
$\min_{i,g}c_i(g)>0$, set
$\widehat C_i(g)=c_i(g)$. Otherwise, the algorithm scales each agent's cost vector to take integer values.
Specifically, we express each cost as an
irreducible fraction
$c_i(g)=u_{ig}/d_{ig}$, where
$u_{ig}\in\mathbb Z_{\ge0}$ and zero is written
as $0/1$.
For each agent $i$, let
$D_i=\operatorname{lcm}_{g\in M}d_{ig}$
and define the scaled integer costs
\(
C_i(g)=D_i c_i(g)
\)
and positive working costs
\(
\widehat C_i(g)=C_i(g)+\varepsilon/(32m)
\). All subsequent stages use
$\widehat C_i$.
The general branch uses the following parameters.
\begin{equation}\label{eq:parameters}
t=\frac{\varepsilon}{100},\quad \beta=\frac{1+2t}{2},\quad U=a=1+t, \quad
\rho=1-t-\beta>0.
\end{equation}
Here $\beta$ is the capped income contribution of a single chore, $\rho=1/2-2t$ determines the minimum price scale for rounded bundles, and $a$ controls the approximation when prices are translated back into costs. 
The income bound $U \geq E_i$ governs the capped prices during the rounding procedure. 
Since $m>2n$ on this branch of the algorithm, we have $m\beta>n$, ensuring sufficient total capped income. The three stages are detailed below, with complete proofs in Appendix~\ref{app:main-proofs}.

\begin{algorithm}[htbp]

\caption{$(3+\varepsilon)$-EFX allocation algorithm}\label{alg:main}

\normalsize

\begin{algorithmic}[1]

\Require A cost matrix $(c_i(g))_{i\in N, g\in M}$ where each $c_i(g) \in \mathbb{Q}$; $0< \varepsilon \leq 1$

\Ensure A complete $(3+\varepsilon)$-EFX allocation


\If{$m\leq 2n$}
    \State \Return $\BRBI(c,1)$
    \Comment{Section~\ref{sec:brbi}}
\EndIf

\If{$\min_{i,g}c_i(g)>0$}
    \State $\widehat C_i(g)\gets c_i(g)$ for all $i,g$
\Else
    \State Scale each row of the cost matrix to obtain
    integer costs $C_i(g)$
    \State $\widehat C_i(g)\gets
    C_i(g)+\varepsilon/(32m)$ for all $i,g$
\EndIf

\State For every bundle $S\subseteq M$, define
$\widehat C_i(S)=\sum_{g\in S}\widehat C_i(g)$

\State Set $t,\beta,U,a,\rho$ as in Eq.~\eqref{eq:parameters}

\State Compute $(p,\gamma,w)\gets\MFCP(\widehat C,t,\beta)$ using Algorithm~\ref{alg:flow}

\State Compute $Y\gets\SSR(p,\gamma,w)$ using Algorithm~\ref{alg:ssr}

\State Split each $Y_i$ into residual bundles $S_i$ and packets as in Section~\ref{sec:packets}

\State $\lambda_0\gets3+\varepsilon/4$

\State $X\gets\APR(\widehat C,S,K,p,\gamma,\rho,\lambda_0)$ using Algorithm~\ref{alg:repair} \Comment{single-chore packets, $k=1$} 

\State \Return $X$

\end{algorithmic}

\end{algorithm}


\subsection{Max-Flow Capped Price}\label{sec:flow}
In the MFCP stage (Algorithm~\ref{alg:flow}), we compute a fractional solution based on the capped-price structure (Definition~\ref{def:capped-price}). 
To balance capped incomes, we construct a flow network in which the
eligible edges provide the approximate MPB support required by Proposition~\ref{prop:mpb}.

\begin{algorithm}[h]

\caption{Max-Flow Capped Price  ($\MFCP$)}\label{alg:flow}

\normalsize

\begin{algorithmic}[1]

\Require Positive costs $(\widehat{C}_{i}(g))_{i\in N, g\in M}$ produced by Alg.~\ref{alg:main}; $0<t\le1/100$; $0<\beta<1$; $m\beta>n$

\Ensure A capped-price structure $(p,\gamma,w)$ satisfying Definition~\ref{def:capped-price}

\State Set $v_i\gets1$ for every agent $i\in N$

\Loop

  \State Set $r_g(v)\gets\min_{i\in N}\{ v_i\cdot \widehat{C}_i(g)\}$ for every chore $g\in M$

  \State Compute $s,p,q$ using Eq.~\eqref{eq:water}

  \State Compute a maximum flow $f$ in the network~\eqref{eq:network}; let its value be $F$

  \If{$F=n$}

    \State Set $w_{ig}\gets f_{ig}/q_g$ and $\gamma_i\gets\min_{g\in M} \{\widehat{C}_i(g)/p_g\}$

    \State \Return $(p,\gamma,w)$

  \EndIf

\State Let $S$ be the set of agents reachable from $s_0$ in the residual network

  \State For every $i\in S$, update $v_i\gets(1+t)\cdot v_i$

\EndLoop

\end{algorithmic}

\end{algorithm}



In MFCP, all agent weights are initially set to $v_i = 1$, and the base price is defined as $r_g(v)=\min_{i\in N}v_i\widehat C_i(g)$.
This base price corresponds to the minimum weighted cost, ensuring that every chore has at least one eligible recipient.
We construct a common water level $s>0$ and scale these prices to ensure the total capped income equals $n$:
\begin{equation}\label{eq:water}
\left(\sum_{g\in M}\min\{s\cdot r_g(v),\beta\}\right) = n; \qquad p_g = s\cdot r_g(v), \qquad q_g = \min\{ p_g, \beta \}.
\end{equation}
The condition $m\beta>n$ guarantees that the first equation has a unique solution: the left-hand side increases continuously from zero and crosses $n$ before all chores are capped. 
By sorting the base prices to determine the prefix of $h$ capped chores, the water level can be exactly computed as $s= (n-h\cdot \beta)/ (\sum_{g: p_g <\beta} r_g(v))$. 
Furthermore, for any fractional allocation $w$, the capped incomes satisfy
$\sum_{i\in N} E_i = \sum_{g\in M}q_g\sum_{i\in N} w_{ig} = n$. 
Thus, 
normalizing total capped income to $n$ yields a target income approximately $1$ per agent, reducing the fractional fairness requirement to  capacity constraints in a flow network.

\paragraph{Flow Network} 
We construct a network which has a source $s_0$, chore nodes, agent nodes, and a sink $t_0$. Its edges and capacities are:
\begin{equation}\label{eq:network}
\begin{aligned}
s_0\longrightarrow g &: q_g, &  g\in M;\\
g\longrightarrow i~ &: n, & \text{If }v_i\widehat C_i(g)\le(1+t)r_g(v); \\
~i\longrightarrow t_0 &: 1+t/n, &  i\in N.
\end{aligned}
\end{equation}
The source-to-chore capacity equals the chore's total capped income contribution. Each agent-to-sink capacity is slightly above $1$, preventing any agent from receiving too much income. 
A chore-to-agent edge exists only when the agent's weighted cost is within a factor $1+t$ of its minimum.

\begin{figure}[htbp]  
    \centering
    \includegraphics[width=0.9\linewidth]
    {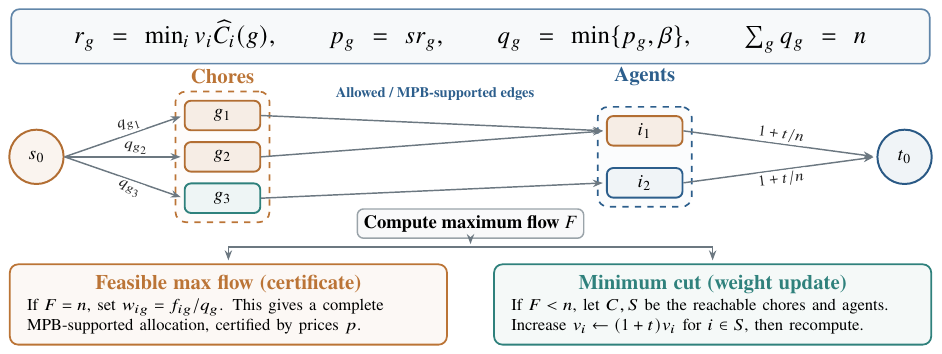}
    \caption{Flow network and weight update in MFCP.}
    \label{fig:mfcp}
\end{figure}

Starting with $v_i=1$, MFCP repeatedly computes the base prices, water level, and a maximum flow in this network.
If the flow value $F$ reaches $n$ (Lines 6-8), every source-to-chore edge is saturated. 
All capped income is sent, and $w_{ig}=f_{ig}/q_g$ defines a complete fractional allocation that preserves approximate MPB support.
Otherwise, if $F < n$ (Lines 10-11), a minimum cut of the residual network determines which agent weights to increase before prices and flows are recomputed. The algorithm continues until a full flow is found; the cut-progress lemma and potential bound below guarantee termination.
Figure~\ref{fig:mfcp} illustrates the MFCP flow network and how the residual cut determines the agent-weight updates.
\begin{example}\label{example}
Consider an instance with two agents and five chores. For agent 1, her costs are: $c_1(g_1) = c_1(g_2) = c_1(g_3) = 3$ and $c_1(g_4) = c_1(g_5) = 1$. For agent 2, $c_2(g_j) = 1.02\cdot c_1(g_j)$ for every $g_j$.
Set $t =0.01$. By Eq.(\ref{eq:parameters}), $\beta = 0.51$, $U = 1.01$, 
$\rho = 0.48$.
\end{example}


\begin{lemma}\label{lem:full-flow}

If the maximum-flow value $F=n$, Algorithm~\ref{alg:flow} outputs a capped-price structure $(p,\gamma,w)$ satisfying Definition~\ref{def:capped-price}, with the stronger income bounds
\[
1-\frac{(n-1)\cdot t}{n}\le E_i\le1+\frac{t}{n}, \quad \forall i\in N.
\]
\end{lemma}

\begin{lemma}[Progress from cut updates]\label{lem:cut}
Suppose  $F<n$, and let $C$ and $S$ be the sets of chores and agents reachable from $s_0$ in the residual network. We have $C \neq \emptyset$, $S\neq \emptyset$, and
\begin{equation}\label{eq:cut-excess}
q(C)>(1+t/n)\cdot|S|.
\end{equation}
The updated base prices satisfy that
$r'_g=(1+t)\cdot r_g$ for any $g\in C$;  $r'_g\ge r_g$ for any $g\notin C$.
\end{lemma}

MFCP must eventually find a full flow of value $n$ and return a structure. Indeed, every unsuccessful flow yields the strict cut inequality~\eqref{eq:cut-excess}, so the update increases the potential defined in the appendix by a positive amount. As the potential has a uniform upper bound, the algorithm must terminate after finitely many updates. Theorem~\ref{thm:flow} gives the polynomial iteration bound.


\begin{theorem}\label{thm:flow}
Let $\widehat C_{\max}=\max_{i,g}\widehat{C}_i(g)$ and $\widehat C_{\min}=\min_{i,g}\widehat{C}_i(g)>0$. The number of times Algorithm~\ref{alg:flow}
invokes a maximum-flow algorithm is at most
\begin{equation}\label{eq:flow-calls}
1+\left\lceil\frac{n^2\log(\widehat C_{\max}/\widehat C_{\min})}{t\log(1+t)}\right\rceil.
\end{equation}
\end{theorem}

Recall that $L_c$ is the total binary encoding
length of the input cost matrix. By construction,
the working costs satisfy
\(
\widehat C_{\max}\le 2^{2L_c+1},
\widehat C_{\min}\ge
\frac{\varepsilon}{32m}2^{-L_c}.
\)
Indeed, in the scaling case,
\(
D_i=\operatorname{lcm}_{g\in M}d_{ig}
\le\prod_{g\in M}d_{ig}\le 2^{L_c},
\)
and hence
\(
C_i(g)=\frac{D_i}{d_{ig}}u_{ig}
\le 2^{2L_c}.
\)
In the unscaled case, each positive input cost
is at most $2^{L_c}$ and at least $2^{-L_c}$.
Moreover, in the scaling case, the scaled costs
are nonnegative integers and at least one is
zero, so $\widehat C_{\min}=\varepsilon/(32m)$.
Thus, the above bounds hold in both cases.
Consequently,
\(
\log\frac{\widehat C_{\max}}{\widehat C_{\min}}
\le
\log\left(\frac{64m}{\varepsilon}2^{3L_c}\right)
=O\left(L_c+\log m+\log(1/\varepsilon)\right).
\)
Since $t=\varepsilon/100$, Algorithm~\ref{alg:flow} runs in time polynomial in
$n$, $m$, $L_c$, and $1/\varepsilon$.

\paragraph{Running Time of MFCP.}
Each iteration builds a flow network with $n+m+2$ vertices and $O(nm+n+m)$ edges. Edmonds--Karp requires $O((n+m)(nm+n+m)^2)$ arithmetic operations per iteration. By Theorem~\ref{thm:flow}, the total arithmetic complexity is
\[
O\left(
\left(
1+\frac{n^2}{\varepsilon^2}
\left(L_c+\log m+\log\frac1\varepsilon\right)
\right)
(n+m)(nm+n+m)^2
\right).
\]

We next round the fractional shares while preserving the cost-price relation and controlling the loss in capped income. This motivates sorted-slot rounding over independent randomized rounding.

\subsection{Rounding and Packet Decomposition}\label{sec:rounding}\label{sec:packets}

\paragraph{Sorted-Slot Rounding.}

For each agent $i$, create $\ell_i=\lceil\sum_gw_{ig}\rceil$ unit-capacity slots and place her fractional shares into them in nonincreasing order of $q_g$. Earlier slots therefore encounter larger chores. Since each slot is matched to at most one chore, the first slot can lose at most $\beta$ in capped income; later slots can be compared with adjacent fractional loads. This ordering provides both a lower bound on rounded income and an upper bound on income after removing the largest chore.

Among chores with $q_g=\beta$, break ties by placing chores with $p_g>\beta$ first. Although their capped prices are equal, this tie-breaking rule ensures that high-price chores can only touch the early slots, controlling their number as well as their total capped price.

\begin{algorithm}[htbp]

\caption{Sorted-Slot Rounding (\SSR)}\label{alg:ssr}

\normalsize

\begin{algorithmic}[1]

\Require Certificate $(p,\gamma,w)$; $q_g=\min\{p_g,\beta\}$

\Ensure Complete integral allocation $Y$

\For{each agent $i$}

\State Create $\ell_i=\lceil\sum_gw_{ig}\rceil$ unit slots

\State Place shares in nonincreasing order of $q_g$, breaking ties in favor of high-price chores

\EndFor

\State Set $L\gets\sum_i\ell_i$ and add $L-m$ dummy chores of $q$-size zero

\State Fill the remaining slot capacity with dummy-chore shares

\State Build the chore--slot support graph using positive shares

\State Find a perfect matching in the support graph

\State Assign each real chore to the agent owning its matched slot

\State Remove the dummy chores and \Return $Y$

\end{algorithmic}

\end{algorithm}

Some slots may be only partially filled by real chore shares. After adding zero-price dummy chores to fill the remaining capacity, every chore has total fractional assignment $1$ and every slot receives total fractional assignment $1$. Construct the bipartite support graph using the positive chore--slot shares. For any set $T$ of chores, its total fractional mass $|T|$ must be assigned to neighboring slots; as each slot has capacity $1$, there must be at least $|T|$ such slots. This verifies Hall's condition, so the support graph has an integral perfect matching. Each real chore is matched only to a slot of an agent who originally held a positive share of that chore, preserving approximate MPB support. We finally discard the dummy chores.
We illustrate the execution of MFCP and SSR on the
instance in Example~\ref{example}, as shown in
Figures~\ref{fig:mf-acp} and~\ref{fig:ssr-steps}, respectively.
\begin{figure}[]
\begin{minipage}{0.49\textwidth}
\centering
\begingroup
\definecolor{ssrone}{RGB}{205,224,244}
\definecolor{ssrtwo}{RGB}{207,234,225}
\definecolor{ssrthree}{RGB}{247,229,189}
\definecolor{ssrfour}{RGB}{245,211,202}
\definecolor{ssrfive}{RGB}{224,213,239}
\begin{tikzpicture}[x=1.30cm,y=1cm,>=Stealth,
  every node/.style={font=\small},
  terminal/.style={circle,draw=black!60,fill=black!3,minimum size=6mm,inner sep=2pt},
  chore/.style={circle,draw=black!60,minimum size=6mm,inner sep=1pt},
  agent/.style={circle,draw=black!60,fill=white,minimum size=7mm,inner sep=1pt},
  arc/.style={->,draw=black!65,line width=.55pt},
  added/.style={->,draw=blue!65!black,dashed,line width=.7pt}]
\node[terminal] (s) at (0,0) {$s_0$};
\node[chore,fill=ssrone] (g1) at (2,2) {$g_1$};
\node[chore,fill=ssrtwo] (g2) at (2,1) {$g_2$};
\node[chore,fill=ssrthree] (g3) at (2,0) {$g_3$};
\node[chore,fill=ssrfour] (g4) at (2,-1) {$g_4$};
\node[chore,fill=ssrfive] (g5) at (2,-2) {$g_5$};
\node[agent] (a1) at (3.6,1.4) {$1$};
\node[agent] (a2) at (3.6,-1.4) {$2$};
\node[terminal] (z) at (5,0) {$t_0$};
\foreach \g/\q in {g1/0.51,g2/0.51,g3/0.51,g4/0.235,g5/0.235}{
  \draw[arc] (s)--(\g) node[pos=.72,sloped,above=2pt,inner sep=1pt] {$\q$};
}
\foreach \g in {g1,g2,g3,g4,g5}{\draw[arc] (\g)--(a1);}
\foreach \g in {g1,g2,g3,g4,g5}{\draw[added] (\g)--(a2);}
\draw[arc] (a1)--(z) node[pos=.48,above right,inner sep=1pt] {$1.005$};
\draw[arc] (a2)--(z) node[pos=.48,below right,inner sep=1pt] {$1.005$};
\draw[arc] (-0.7,-2.7)--(-0.1,-2.7);
\draw[added] (2.7,-2.7)--(3.3,-2.7);
\node[anchor=west,font=\footnotesize] at (-0.1,-2.7)
  {present in both rounds};
\node[anchor=west,font=\footnotesize] at (3.3,-2.7)
  {admitted in round $2$};
\end{tikzpicture}
\caption{\footnotesize Execution of Alg.~\ref{alg:flow} on Example~\ref{example}. In round 1, $v = (1,1)$, $F = 1.005$; in  
round 2, $v = (1.01, 1)$, $F = 2$.}
\label{fig:mf-acp}
\endgroup
\end{minipage}
\begin{minipage}{0.49\textwidth}
\centering
\begingroup
\definecolor{ssrone}{RGB}{205,224,244}
\definecolor{ssrtwo}{RGB}{207,234,225}
\definecolor{ssrthree}{RGB}{247,229,189}
\definecolor{ssrfour}{RGB}{245,211,202}
\definecolor{ssrfive}{RGB}{224,213,239}
\begin{tikzpicture}[x=1cm,y=1cm,every node/.style={font=\footnotesize}]
\foreach \x/\s in {1.15/1,3.4/2,5.65/3}{
  \node at (\x+1,.47) {slot $\s$};
}
\node[anchor=east] at (1.0,0) {agent $1$};
\node[anchor=east] at (1.0,-.60) {agent $2$};
\fill[ssrone] (1.15,-.24) rectangle (2.65,.24);
\fill[ssrtwo] (2.65,-.24) rectangle (3.15,.24);
\fill[ssrtwo] (3.4,-.24) rectangle (3.9,.24);
\fill[ssrthree] (3.9,-.24) rectangle (4.4,.24);
\fill[ssrfour] (4.4,-.24) rectangle (5.4,.24);
\fill[ssrone] (1.15,-.84) rectangle (1.65,-.36);
\fill[ssrtwo] (1.65,-.84) rectangle (2.65,-.36);
\fill[ssrthree] (2.65,-.84) rectangle (3.15,-.36);
\fill[ssrthree] (3.4,-.84) rectangle (4.4,-.36);
\fill[ssrfour] (4.4,-.84) rectangle (5.4,-.36);
\foreach \y/\split/\gmid/\dmid in {0/6.65/6.15/7.15,-.60/6.65/6.15/7.15}{
  \fill[ssrfive] (5.65,\y-.24) rectangle (\split,\y+.24);
  \fill[black!5] (\split,\y-.24) rectangle (7.65,\y+.24);
  \fill[pattern=north east lines,pattern color=black!25]
    (\split,\y-.24) rectangle (7.65,\y+.24);
  \foreach \x in {1.15,3.4,5.65}{
    \draw[draw=black!65] (\x,\y-.24) rectangle (\x+2,\y+.24);
  }
  \draw[draw=black!45] (\split,\y-.24)--(\split,\y+.24);
  \node at (\gmid,\y) {$g_5$};
  \node[fill=white,inner sep=.5pt] at (\dmid,\y) {$d$};
}
\draw[draw=black!45] (2.65,-.24)--(2.65,.24);
\foreach \x in {3.9,4.4}{\draw[draw=black!45] (\x,-.24)--(\x,.24);}
\foreach \x in {1.65,2.65}{\draw[draw=black!45] (\x,-.84)--(\x,-.36);}
\draw[draw=black!45] (4.4,-.84)--(4.4,-.36);
\node at (1.90,0) {$g_1$};\node at (2.90,0) {$g_2$};
\node at (3.65,0) {$g_2$};\node at (4.15,0) {$g_3$};\node at (4.90,0) {$g_4$};
\node at (1.40,-.60) {$g_1$};\node at (2.15,-.60) {$g_2$};\node at (2.90,-.60) {$g_3$};
\node at (3.90,-.60) {$g_3$};\node at (4.90,-.60) {$g_4$};
\begin{scope}[yshift=-1.37cm,
  item/.style={circle,draw=black!60,minimum size=5.5mm,inner sep=1pt},
  slot/.style={rectangle,rounded corners=2pt,draw=black!60,minimum width=10mm,minimum height=5.5mm,inner sep=1pt},
  support/.style={draw=black!35,dashed,line width=.5pt},
  matching/.style={draw=blue!65!black,line width=1.1pt}]
\node[anchor=east] at (.58,-.35) {chores};
\node[anchor=east] at (.58,-2.10) {slots};
\node[item,fill=ssrone] (g1) at (1,-.35) {$g_1$};
\node[item,fill=ssrtwo] (g2) at (2.25,-.35) {$g_2$};
\node[item,fill=ssrthree] (g3) at (3.5,-.35) {$g_3$};
\node[item,fill=ssrfour] (g4) at (4.75,-.35) {$g_4$};
\node[item,fill=ssrfive] (g5) at (6,-.35) {$g_5$};
\node[item,fill=black!5] (gd) at (7.25,-.35) {$d$};
\node[slot] (s11) at (1,-2.10) {$(1,1)$};
\node[slot] (s12) at (2.25,-2.10) {$(1,2)$};
\node[slot] (s21) at (3.5,-2.10) {$(2,1)$};
\node[slot] (s22) at (4.75,-2.10) {$(2,2)$};
\node[slot] (s13) at (6,-2.10) {$(1,3)$};
\node[slot] (s23) at (7.25,-2.10) {$(2,3)$};
\draw[support] (g1)--(s21);
\draw[support] (g2)--(s11);
\draw[support] (g2)--(s21);
\draw[support] (g3)--(s12);
\draw[support] (g3)--(s22);
\draw[support] (g4)--(s12);
\draw[support] (g5)--(s23);
\draw[support] (gd)--(s13);
\draw[matching] (g1)--(s11);
\draw[matching] (g2)--(s12);
\draw[matching] (g3)--(s21);
\draw[matching] (g4)--(s22);
\draw[matching] (g5)--(s13);
\draw[matching] (gd)--(s23);
\draw[matching] (.75,-2.90)--(1.35,-2.90);
\node[anchor=west] at (1.50,-2.90) {matching edges};
\draw[support] (4.1,-2.90)--(4.7,-2.90);
\node[anchor=west] at (4.7,-2.90) {other support edges};
\end{scope}
\end{tikzpicture}
\endgroup

\caption{\footnotesize Execution of Alg.~\ref{alg:ssr} on Example~\ref{example}. Each quarter-slot has mass $1/4$; the dummy has $q_d = 0$.}
\label{fig:ssr-steps}

\end{minipage}
\end{figure}

\begin{theorem}\label{thm:ssr}

Let $H=\{g:p_g>\beta\}$. Algorithm~\ref{alg:ssr} runs in $O(nm\log m+(m+n)^{5/2})$ time, returns a complete allocation, and satisfies
\begin{align}
q(Y_i)&\ge E_i-\beta,\label{eq:ssr-low}\\
q(Y_i)-\max_{g\in Y_i}q_g&\le E_i,\label{eq:ssr-tail}\\
|Y_i\cap H|&\le\left\lceil\sum_{g\in H}w_{ig}\right\rceil.\label{eq:ssr-high}
\end{align}
All ownership edges for real chores retain the approximate MPB guarantee.

\end{theorem}

Substituting the parameters from~\eqref{eq:parameters} yields
\begin{equation}\label{eq:rounded-interface}
p(Y_i)\ge\rho,\quad q(Y_i)\le U+\beta,\quad |Y_i\cap H|\le k+1.
\end{equation}
By Theorem~\ref{thm:ssr}, the rounded allocation preserves approximate MPB support and contains at most $k+1$ high-price chores per agent.
Moreover, Proposition~\ref{prop:mpb}(1),
applied to $\widehat C_i$, gives
\(
\widehat C_i(S_i)\le a\gamma_i p(S_i),
\widehat C_i(K)\ge\gamma_i\rho
\)
for every residual bundle $S_i$ and every
packet $K$ with $p(K)\ge\rho$.
Since the capped-price guarantee does not imply pEF1 under the original prices, we decompose the rounded bundles into packets for reallocation.

\paragraph{Packet Decomposition} 
By Theorem~\ref{thm:ssr}, each rounded bundle contains at most $k+1$ high-price chores.
We use this bound to extract at most two packets, each containing at most $k$ chores and having price greater than $\rho$, and place all remaining chores in a residual bundle $S_i$.
A recipient holding only a packet has zero EF$k$X residual, since the packet contains at most $k$ chores. 
This allows APR to transfer packets as whole units.

Formally, a packet is a set of at most $k$ chores that is transferred as a unit. Decompose each integral bundle as
\[
Y_i=S_i\mathbin{\dot\cup}K_{i,1}\mathbin{\dot\cup}\cdots\mathbin{\dot\cup}K_{i,\nu_i},
\qquad \nu_i\in\{0,1,2\}.
\]
For EFX, every packet contains one chore with price greater than $\beta>\rho$.

For $k\ge2$, first consider the set $T_i$ of the $k$ highest-price chores. If the $(k+1)$-st chore also has price greater than $\rho$, make $T_i$ and that single chore separate packets, covering the possible $k+1$ high-price chores. If the next chore is not that expensive, create one packet only when $p(T_i)>\rho$; otherwise create no packet. The number of extracted packets, $\nu_i\in\{0,1,2\}$, determines the upper bound on $p(S_i)$ given in Table~\ref{tab:budgets}. In Section~\ref{sec:factor}, we combine these bounds with Proposition~\ref{prop:mpb}(1) to establish the residual-cost condition $\widehat C_i(S_i)\le(\lambda-\nu_i)\gamma_i\rho$ required by Lemma~\ref{lem:repair}.

\begin{lemma}[Residual-bundle budgets]\label{lem:packet-budget}

Each packet contains at most $k$ chores and has price greater than $\rho$. If $\nu_i=0$, then $p(S_i)\ge\rho$. 
The upper bounds on $p(S_i)$ for each case are listed in Table~\ref{tab:budgets}. By Proposition~\ref{prop:mpb}(1), these bounds imply the corresponding upper bounds on $\widehat C_i(S_i)$.
\end{lemma}

\begin{table}[t]

\caption{Upper bounds on residual prices $p(S_i)$.}\label{tab:budgets}

\centering

\begin{tabular}{@{}cccc@{}}

\toprule

 & $\nu_i=0$ & $\nu_i=1$ & $\nu_i=2$\\

\midrule

$k=1$ & $U+\beta$ & $U$ & $U-\beta$\\

$k\ge2$ & $U+\rho$ & $U$ & $U-k\beta$\\

\bottomrule

\end{tabular}

\end{table}
The reallocation stage relies only on these bounds and does not require the final allocation to retain MPB support.

\subsection{Reallocation}\label{sec:repair}

Rounding and packet decomposition bound each agent's residual cost, but may not eliminate envy toward a particularly cheap comparison bundle. APR uses two passes in the same fixed order. In the first pass, each active agent chooses the $\nu_i$ cheapest packets available to her, denoted $Q_{i,1},\ldots,Q_{i,\nu_i}$, from a common pool. She keeps $S_i$ and combines it with the selected packets to form a temporary complete allocation. In the second pass, agents are checked and repaired in the same order.

The fixed selection order ensures that, for each active agent $i$, every packet selected later is at least as costly to $i$ as the most expensive packet she selected. This property provides a lower bound on the cost of later-selected packets in subsequent comparisons. A recipient of a pure packet has at most $k$ chores and hence satisfies exact EF$k$X; a composite bundle contains an earlier complete bundle and therefore cannot become cheaper to any observer. Together, these properties prevent later exchanges from breaking fairness already established.

\begin{lemma}[Adaptive packet reallocation]
\label{lem:repair}
Suppose $\lambda\ge2$. Each agent has at most two packets, each packet contains at most $k$ chores and has price at least $\rho>0$. Assume that
\begin{align}
p(S_i)\ge\rho,\quad \widehat C_i(S_i)
&\le\lambda\gamma_i\rho
&&(\nu_i=0),\label{eq:repair-zero}\\
\widehat C_i(S_i)
&\le(\lambda-\nu_i)\gamma_i\rho
&&(\nu_i\in\{1,2\}).\label{eq:repair-active}
\end{align}
Then Algorithm~\ref{alg:repair} runs in $O(n^2m+nm\log m)$ time and returns a complete $\lambda$-EF$k$X allocation.
\end{lemma}

\begin{algorithm}[t]

\caption{Adaptive Packet Reallocation($\APR$)}\label{alg:repair}

\normalsize

\begin{algorithmic}[1]

\Require A packet decomposition satisfying Lemma~\ref{lem:repair}

\Ensure Complete $\lambda$-EF$k$X allocation

\State Each agent $i$ with $\nu_i>0$ chooses her
$\nu_i$ cheapest available packets in a fixed order

\State Set $X_i\gets S_i\cup\bigcup_\ell Q_{i,\ell}$ for every agent $i$

\For{active agent $i$, in the same order}

  \State Let $B_1=X_{j_1}$ be the bundle cheapest to agent $i$ among the other bundles

  \State $b_1\gets \widehat C_i(B_1)$

  \If{$R_i^k(X_i)\le\lambda b_1$}

    \State \textbf{continue}

  \EndIf

  \If{$\nu_i=1$}

    \State $(X_i,X_{j_1})\gets(S_i\cup B_1,Q_{i,1})$

  \Else

\State Denote the two packets by $Q^+,Q^-$, ordered so that $\widehat C_i(Q^+)\ge \widehat C_i(Q^-)$

    \State Let $B_2=X_{j_2}$ be the second-cheapest bundle for agent $i$

    \State $b_2\gets \widehat C_i(B_2)$; if no such bundle exists, set $b_2=+\infty$

    \If{$\widehat C_i(Q^-)\le \widehat C_i(B_2)$}

      \State $(X_i,X_{j_1})\gets(S_i\cup Q^-\cup B_1,Q^+)$

    \Else

      \State $X_i\gets S_i\cup B_1\cup B_2$

      \State $(X_{j_1},X_{j_2})\gets(Q^+,Q^-)$

    \EndIf

  \EndIf

\EndFor

\State \Return $X$

\end{algorithmic}

\end{algorithm}

Algorithm~\ref{alg:repair} chooses an exchange according to the number of packets held by the active agent. If the agent holds one packet, a violation is repaired by exchanging that packet for the cheapest comparison bundle $B_1$; the agent keeps $S_i$ and receives $B_1$, while the other agent receives a pure packet of at most $k$ chores. If the agent holds two packets, she compares the cheaper packet $Q^-$ with the second-cheapest comparison bundle $B_2$. If $\widehat C_i(Q^-)\le \widehat C_i(B_2)$, retain $Q^-$, send out the more expensive packet, and receive $B_1$; if $\widehat C_i(Q^-)>\widehat C_i(B_2)$, send out both packets and receive $B_1,B_2$. These cases bound the repaired agent's own cost by the cheapest  comparison cost and avoid creating a new, cheaper comparison bundle for future agents.

The proof must maintain both the initial bundles of unprocessed agents and the fairness already achieved for processed agents. Lemma~\ref{lem:repair} reduces the required conditions to a lower bound $\rho$ on bundle prices and residual-cost budgets determined by the number of packets. The complete proof is given in Appendix~\ref{app:main-proofs}.
\paragraph{Running Time}
For $m>2n$, combining the running times of MFCP, SSR, and APR, together with the packet decomposition, yields the following total arithmetic complexity:
\[
O\left(
\frac{n^4m^3}{\varepsilon^2}
\left(L_c+\log\frac1\varepsilon\right)
\right).
\]
For $m\le 2n$, Algorithm~\ref{alg:main} applies BRBI,
which runs in polynomial time.
Thus, in both cases, the algorithm runs in time polynomial
in $n$, $m$, $L_c$, and $1/\varepsilon$.
\subsection{Approximation Analysis}\label{sec:main-proof}\label{sec:factor}
The approximation factor is determined by the residual-price bounds in Table~\ref{tab:budgets} and the reallocation guarantee in Lemma~\ref{lem:repair}.
By Theorem~\ref{thm:ssr} and Proposition~\ref{prop:mpb}(1), applied to $\widehat C_i$, the residual bundle $S_i\subseteq Y_i$ satisfies $\widehat C_i(S_i)\le a\gamma_i p(S_i)$, where $a=1+t$.
Moreover, every comparison bundle $B$ with $p(B)\ge\rho$ satisfies $\widehat C_i(B)\ge\gamma_i\rho$.

For EFX, let $\nu_i\in\{0,1,2\}$ denote the number of packets associated with agent $i$. Table~\ref{tab:budgets} gives $p(S_i)\le U+(1-\nu_i)\beta$.
By Proposition~\ref{prop:mpb}(1), it suffices to choose
$\lambda$ such that, for every
$\nu_i\in\{0,1,2\}$,
\(
a\bigl(U+(1-\nu_i)\beta\bigr)
\le(\lambda-\nu_i)\rho.
\)
The smallest such $\lambda$ is therefore
\begin{equation}\label{eq:efx-three-cases}
\max\left\{
\frac{a(U+\beta)}{\rho},
1+\frac{aU}{\rho},
2+\frac{a(U-\beta)}{\rho}
\right\}
=\frac{(1+t)(3/2+2t)}{1/2-2t}
\le 3+20t.
\end{equation}
Here, $0<t\le1/100$. We use $a=U=1+t$, $\beta=1/2+t$, and $\rho=1/2-2t$; the first case dominates.
Choosing $t=\varepsilon/100$ gives a
$(3+\varepsilon/5)$-EFX guarantee with respect to
the working costs $\widehat C_i$.
If all original costs are positive, then
$\widehat C_i=c_i$ and the guarantee applies
directly. Otherwise, the perturbation analysis
in Appendix~\ref{app:main-proofs} transfers
the guarantee to the original costs $c_i$.

\paragraph{EF$k$X Extension.}
For $k\ge2$, choosing $\beta=(1+2t)/(k+1),\rho=1-t-\beta
=[k-(k+3)t]/(k+1)$ and packets of size at most $k$ gives, by the same price and support arguments, $\alpha_k(t)=(1+t)(1+(1+t)/\rho) \le2+1/k+8t$.
Hence, the algorithm achieves a complete $(2+1/k+\varepsilon)$-EF$k$X allocation. The proofs of both guarantees, including the perturbation analysis, are given in Appendix~\ref{app:main-proofs}.

\section{Sparse Instances}\label{sec:brbi}
For instances with relatively few chores, BRBI provides a density-dependent guarantee without requiring a price certificate or matching-based rounding.
\begin{algorithm}[t]
\caption{Balanced Reservation and Block Insertion (\BRBI)}
\label{alg:brbi}
\normalsize
\begin{algorithmic}[1]
\Require Cost functions $(c_i)_{i\in N}$ and integer $k\ge1$
\Ensure A complete allocation $X$

\If{$m\le kn$}
    \State \Return any complete allocation $X$ with $|X_i|\le k$ for all $i$
\EndIf

\State Choose balanced integers $r_1,\ldots,r_n\ge0$ with
$\sum_{i=1}^n r_i=m-kn$ and $|r_i-r_j|\le1$ for all $i,j$
\State $R\gets M$
\For{$i=n,n-1,\ldots,1$}
    \State Let $E_i$ be the $r_i$ cheapest chores in $R$ according to $c_i$
    \State $R\gets R\setminus E_i$
\EndFor

\State Partition $R$ into $n$ base bundles of size $k$
\State Assign the base bundles bijectively to agents

\For{$i=1,2,\ldots,n$}
    \State Let $j\in\argmin_{h\in N}c_i(X_h)$
    \If{$j=i$}
        \State $X_i\gets X_i\cup E_i$
    \Else
        \State $(X_i,X_j)\gets(X_j\cup E_i,X_i)$
    \EndIf
\EndFor

\State \Return $X$
\end{algorithmic}
\end{algorithm}
\begin{theorem}\label{thm:brbi}
Algorithm~\ref{alg:brbi} runs in polynomial time and returns a complete $\alpha_{\mathrm B}(m,n,k)$-EF$k$X allocation, where
\[
\alpha_{\mathrm B}(m,n,k)
=\max\left\{1,\frac{\lceil(m-kn)/n\rceil}{k}\right\}.
\]
In particular, it achieves exact EF$k$X when $m\le2kn$, and a factor of at most $2-1/k$ when $m\le(3k-1)n$. For $k=1$, this gives exact EFX when $m\le2n$.
\end{theorem}

BRBI uses reverse-order reservation and forward-order insertion to control the cost of added chores. Its construction and proof are given in Appendix~\ref{app:brbi-proof}.
When both algorithms are applicable, we can choose the better guarantee between $\alpha_{\mathrm B}(m,n,k)$ and $2+1/k+\varepsilon$.

\section{Conclusion}\label{sec:conclusion}

We present polynomial-time algorithms for approximate EFX and EF$k$X allocations of indivisible chores under nonnegative rational additive costs. Our main algorithm achieves a $(3+\varepsilon)$-EFX guarantee and extends to a $(2+1/k+\varepsilon)$-EF$k$X guarantee, with the leading constant asymptotically tight. For instances with $m\le 2kn$, BRBI achieves exact EF$k$X, while for $m\le(3k-1)n$, it guarantees a factor of at most $2-1/k$. We also establish an EFX existence lower bound of $11/10$ under additive costs.

\printbibliography[heading=bibliography,title={References}]

\appendix

\section{Main-Algorithm Proofs}\label{app:main-proofs}

This appendix proves the price-certificate, rounding, and reallocation results used in Section~\ref{sec:main-algorithm}.

\begin{proof}[Proof of Lemma~\ref{lem:full-flow}]

The total capacity leaving the source is $n$, so every source edge is saturated. Thus $\sum_i f_{ig}=q_g$ and $\sum_iw_{ig}=1$. Sink capacities give $E_i\le1+t/n$. Since total income equals $n$, subtracting the upper bounds for the other agents yields the stated lower bound.

For every $i,g$, $r_g\le v_i\widehat C_i(g)$,
so $\gamma_i\ge1/(sv_i)$.
On every positive-flow edge,
\[
\widehat C_i(g)
\le\frac{(1+t)r_g}{v_i}
=\frac{(1+t)p_g}{sv_i}
\le(1+t)\gamma_i p_g.
\]
The universal lower bound follows from the definition of $\gamma_i$.
\end{proof}

\begin{proof}[Proof of Lemma~\ref{lem:cut}]

The capacity of the reachable cut is $F<n$, so no middle edge of capacity $n$ can cross the cut. Hence all eligible recipients of chores in $C$ belong to $S$, and
\[
F=q(M\setminus C)+(1+t/n)|S|<q(M)=n.
\]
This proves~\eqref{eq:cut-excess} and shows $C\ne\varnothing$. Every chore has at least one eligible recipient, so $S\ne\varnothing$.

For $g\in C$, all minimizers of the old weighted cost belong to $S$, whereas agents outside $S$ have weighted cost strictly greater than $(1+t)r_g$. Increasing the weights in $S$ therefore gives $r'_g=(1+t)r_g$; all other base prices are nondecreasing.
\end{proof}

\begin{proof}[Proof of Theorem~\ref{thm:flow}]

Let $\mathcal Q=\{x\in[0,\beta]^m:\sum_gx_g=n\}$ and define
\begin{equation}\label{eq:potential}
\Psi(v)=\max_{x\in\mathcal Q}\sum_gx_g\log\frac{r_g(v)}{x_g}
-\sum_i\log v_i,
\end{equation}
We use the convention $0\log(r/0)=0$. The capped vector $q$ obtained from the water-level calculation maximizes the first term. The objective is concave; derivatives at uncapped coordinates equal $-\log s-1$, while derivatives at capped coordinates are no smaller, so every feasible first-order change is nonpositive.

After a failed update, insert the previous feasible vector $q$ into the new maximization problem. By Lemma~\ref{lem:cut},
\begin{align*}
\Psi(v')-\Psi(v)
&\ge\sum_gq_g\log(r'_g/r_g)-|S|\log(1+t)\\
&\ge\bigl(q(C)-|S|\bigr)\log(1+t)\\
&>\frac tn\log(1+t).
\end{align*}
The old and new capped prices need not be equal; here we use only the feasibility of the old vector.

Let $v_{\min}=\min_i v_i$. Using $r_g\le v_{\min}\widehat C_{\max}$, $\sum_i\log v_i\ge n\log v_{\min}$, and the entropy bound gives
\(
\Psi(v)\le n\log \widehat C_{\max}+n\log(m/n).
\)
Initially $v=\mathbf 1$, and the uniform vector $x_g=n/m<\beta$ is feasible. Hence
\[
\Psi(\mathbf1)\ge n\log \widehat C_{\min}+n\log(m/n).
\]
The total increase is at most $n\log(\widehat C_{\max}/\widehat C_{\min})$. Dividing by the progress made in each update yields~\eqref{eq:flow-calls}. Since $\log(1+t)\ge t/2$, the number of calls is
\[
O\left(1+\frac{n^2\log(\widehat C_{\max}/\widehat C_{\min})}{t^2}\right).
\]
For bit complexity, after the $j$-th update each weight has the form $v_i=(1+t)^{e_i}$ with $0\le e_i\le j$, so its bit length is polynomially bounded. Base prices, water levels, and capacities are rational numbers of polynomial bit length. Each network has $n+m+2$ vertices and at most $nm+n+m$ edges; after clearing denominators, an exact integer max-flow algorithm such as Edmonds--Karp applies. The returned prices, shares, and $\gamma_i$ values also have polynomial bit length. The potential function is used only in the proof.
\end{proof}

\begin{proof}[Proof of Theorem~\ref{thm:ssr}]

Fix an agent with slots $1,\ldots,\ell$. Let $b_j$ be the fractional $q$-load of slot $j$, and let $v_j$ be the $q$-size of its matched chore (zero for a dummy chore). Sorting implies
\[
v_j\le b_{j-1}\quad(j\ge2),\qquad
v_j\ge b_{j+1}\quad(j<\ell).
\]
This follows because a chore reaching a later slot is no larger than any chore in an earlier full slot; comparing the next slot in the reverse direction gives the corresponding bound. Therefore,
\[
q(Y_i)\ge\sum_{j=2}^{\ell}b_j=E_i-b_1\ge E_i-\beta,
\]
and
\[
q(Y_i)-\max_{g\in Y_i}q_g
\le\sum_{j=2}^{\ell}v_j\le E_i.
\]
Only the first $\lceil\sum_{g\in H}w_{ig}\rceil$ slots can meet high-price fractional shares, proving~\eqref{eq:ssr-high}.

The completed slots and chore shares form a fractional perfect matching, so Hall's condition holds and an integral perfect matching exists. Every real matched edge is supported by the certificate. Finally, $\sum_i\ell_i\le m+n$, so the graph and matching have polynomial size.
\end{proof}

\begin{proof}[Proof of Lemma~\ref{lem:packet-budget}]

When $k=1$, $S_i$ contains only low-price chores, so
\[
p(S_i)=q(Y_i)-\nu_i\beta\le U+(1-\nu_i)\beta.
\]
If no packet is selected, then $S_i=Y_i$,
and the lower bound follows from~\eqref{eq:rounded-interface}.

Now consider the case $k\ge2$. The parameters satisfy
\[
\rho=\frac{k-(k+3)t}{k+1}>\beta,
\qquad
k\beta\ge\rho.
\]
The packet-size and packet-price conditions follow
from the definition. If $\nu_i=2$, the first $k+1$ chores are all high-price chores. The bound on their number implies that all remaining chores are low-price chores. Thus,
\[
p(S_i)=q(Y_i)-(k+1)\beta\le U-k\beta.
\]
If $\nu_i\in\{0,1\}$, we first show $p(Y_i\setminus T_i)\le U$. If there are at most $k$ high-price chores, they all lie in $T_i$, and the remaining chores satisfy $p=q$, so~\eqref{eq:ssr-tail} applies. If there are exactly $k+1$ high-price chores, the cheapest has price at most $\rho$; otherwise the two-packet case would apply. The low-price remainder has price at most $U-k\beta$, hence
\[
p(Y_i\setminus T_i)\le U-k\beta+\rho\le U.
\]
This proves the one-packet budget. If $\nu_i=0$, then $p(T_i)\le\rho$, so $p(S_i)=p(Y_i)\le U+\rho$, while $p(S_i)\ge\rho$. Finally, chores in $S_i$ remain with their rounded owners, so approximate MPB translates the price bounds into cost bounds.
\end{proof}

\begin{proof}[Proof of Lemma~\ref{lem:repair}]
By Proposition~\ref{prop:mpb}(1), applied to
$\widehat C_i$, we have
$\widehat C_i(B)\ge\gamma_i p(B)$
for every agent $i$ and every bundle $B$.
Fix an active agent $i$ and write $d_i=\max_\ell \widehat C_i(Q_{i,\ell})$. Every packet chosen later costs at least $d_i$ to $i$. Each packet has price at least $\rho$, so $d_i\ge\gamma_i\rho$, and initially
\begin{equation}\label{eq:barrier}
\widehat C_i(X_i)\le(\lambda-\nu_i)\gamma_i\rho+\nu_i d_i\le\lambda d_i.
\end{equation}
Every current bundle always has price at least $\rho$. This holds initially by assumption and remains true after an exchange, because a new bundle is either a pure packet or a composite bundle containing an earlier complete bundle. Therefore, for every observer $i$,
\begin{equation}\label{eq:universal-floor}
\widehat C_i(B)\ge\gamma_i\rho\qquad\text{for every current bundle }B.
\end{equation}
\paragraph{Unprocessed agents.}
An exchange never takes away the standard bundle of a future active agent. If $B_1$ is such a bundle, it contains one packet costing at least $d_i$ to agent $i$. The barrier inequality~\eqref{eq:barrier} then implies $R_i^k(X_i)\le\lambda \widehat C_i(B_1)$, so no repair is needed. The second bundle is selected only when $\widehat C_i(Q^-)>b_2$; if $B_2$ belonged to a future active agent, then $b_2\ge d_i\ge \widehat C_i(Q^-)$, a contradiction. Inductively, every active agent retains her standard bundle until her turn.

\paragraph{One-packet case.}
Let $s=\widehat C_i(S_i)$ and $u=\widehat C_i(Q_{i,1})$. By~\eqref{eq:repair-active} and~\eqref{eq:universal-floor}, $s\le(\lambda-1)b_1$. A repair is triggered only if $u>b_1$, since otherwise $s+u\le\lambda b_1$. After the exchange, agent $i$'s total cost is at most $s+b_1\le\lambda b_1$; every comparison bundle still costs at least $b_1$, and the new pure packet costs $u>b_1$. Its recipient holds at most $k$ chores, so the residual cost is zero.

\paragraph{Two-packet case.}
Let $u=\widehat C_i(Q^+)\ge v=\widehat C_i(Q^-)$ and $s=\widehat C_i(S_i)\le(\lambda-2)\gamma_i\rho$. A violation again implies $u>b_1$, because otherwise both packets would cost at most $b_1$ and the total cost would be at most $\lambda b_1$.

If $v\le b_2$, let $\mu=\min\{u,b_2\}$, so $b_1,v,\gamma_i\rho\le\mu$. After the single exchange,
\[
\widehat C_i(X_i)=s+v+b_1\le\lambda\mu.
\]
The new pure packet costs $u$, while every unchanged comparison bundle costs at least $b_2$; hence all comparison costs are at least $\mu$. When $n=2$, the argument remains valid with $b_2=+\infty$.

If $v>b_2$, both packets are sent out, and the new bundle of agent $i$ satisfies
\[
\widehat C_i(X_i)=s+b_1+b_2\le\lambda b_2.
\]
Both new pure packets cost more than $b_2$, and all other comparison bundles already cost at least $b_2$. Thus both branches complete the repair.

\paragraph{Processed agents.}
After repair, $\widehat C_i(X_i)\le\lambda d_i$. If agent \(i\) has one packet, then $b_1<u=d_i$. If she has two packets and only one is exchanged, then $\mu\le u=d_i$. If both packets are exchanged, then \(b_2<v\le d_i\). An agent who is not repaired retains~\eqref{eq:barrier}. Later composite bundles contain an earlier complete bundle, and later pure packets cost at least $d_i$ to agent $i$, so established comparisons persist. If an agent's own bundle is taken, she receives a pure packet and has zero residual cost. Agents with no packets remain fair by~\eqref{eq:repair-zero} and~\eqref{eq:universal-floor}, or become exactly fair after receiving a pure packet.

Every update preserves a complete partition. Each active agent is repaired at most once, and there are at most $2n$ packets; packet selection, cost comparisons, and fairness checks all take polynomial time.
\end{proof}

\begin{lemma}[Approximation-factor budgets]\label{lem:factor} The decomposition in Lemma~\ref{lem:packet-budget} satisfies the hypotheses of Lemma~\ref{lem:repair} with
\[\Lambda_k(t)=\begin{cases}\displaystyle\frac{(1+t)(3/2+2t)}{1/2-2t},&k=1,\\[6pt]\displaystyle(1+t)\left(1+\frac{1+t}{\rho}\right),&k\ge2.\end{cases}\]
For $0<t\le1/100$, $\Lambda_1(t)\le3+20t$ and $\Lambda_k(t)\le2+1/k+8t$ for $k\ge2$.\end{lemma}

\begin{proof}[Proof of Lemma~\ref{lem:factor}]

For $k\ge2$, the zero-packet cost is at most $a\gamma_i(U+\rho)=\Lambda_k(t)\gamma_i\rho$. For the one-packet case, $aU\le(\Lambda_k(t)-1)\rho$; for the two-packet case, $k\beta\ge\rho$ and $a\ge1$ imply $a(U-k\beta)\le(\Lambda_k(t)-2)\rho$.

For $k=1$, $\Lambda_1(t)\rho=a(U+\beta)$ and $a\beta\ge\rho$, so the three columns of Table~\ref{tab:budgets} verify the required bounds. Expanding gives
\begin{align*}
\Lambda_1(t)-3&=\frac{19t+4t^2}{1-4t},\\
\Lambda_k(t)-\left(2+\frac1k\right)
&=\frac{t(4k^2+6k+3-2kt)}{k\bigl(k-(k+3)t\bigr)}\quad(k\ge2).
\end{align*}
For $t\le1/100$, the two expressions are bounded by $20t$ and $8t$, respectively.
\end{proof}

\begin{proof}[Proof of Theorem~\ref{thm:main}
and Corollary~\ref{cor:main-generalization}]
The case $n=1$ is immediate. If $m\le2kn$, Theorem~\ref{thm:brbi} gives exact EF$k$X. Otherwise, if $\min_{i,g}c_i(g)>0$, set
$\widehat C_i=c_i$. If some original cost is zero,
perform row-wise integer scaling and set
\(
C_i(g)=D_i c_i(g),
\widehat C_i(g)=C_i(g)+\xi,
\xi=\frac{\varepsilon}{32m}.
\)
Row-wise scaling preserves all EF$k$X inequalities. The perturbed working costs are strictly positive, and the parameters satisfy $m\beta>n$, so MFCP and SSR apply.
Taking $t=\varepsilon/100$, Lemma~\ref{lem:factor} gives
\[
\Lambda_k(t)\le
\begin{cases}
3+\varepsilon/5,&k=1,\\
2+1/k+2\varepsilon/25,&k\ge2,
\end{cases}
\]
The factor is strictly below $\lambda_0=2+1/k+\varepsilon/4$. Lemma~\ref{lem:repair} therefore gives a $\lambda_0$-EF$k$X allocation $X$ for the working costs $\widehat C$.

If all original costs are positive, then
$\widehat C_i=c_i$, and the guarantee already
holds for the original costs. Otherwise, adding the same $\xi$ to every chore does not
change the ordering within a row. Hence, for
every bundle $B$,
\[
R_{i,\widehat C}^k(B)
=
R_{i,C}^k(B)
+\xi\max\{|B|-k,0\}.
\]
Thus, for every $i,j\in N$,
$$
R_{i,C}^k(X_i)
\le R_{i,\widehat C}^k(X_i)
\le\lambda_0\widehat C_i(X_j)
=\lambda_0 C_i(X_j)
+\lambda_0\xi|X_j|
\le\lambda_0 C_i(X_j)
+\lambda_0\xi m.
$$
Since $\lambda_0\le13/4$, we have
\(
\lambda_0\xi m
\le 13\varepsilon/128
<3\varepsilon/4.
\)
If $C_i(X_j)=0$, then
$R_{i,C}^k(X_i)$ is a nonnegative integer
strictly below $1$, and hence equals zero.
If $C_i(X_j)\ge1$, the remaining multiplicative
slack absorbs the additive perturbation.
Consequently,
\[
R_{i,C}^k(X_i)
\le
\left(2+\frac1k+\varepsilon\right)C_i(X_j).
\]
Since $C_i=D_i c_i$, dividing by $D_i$
transfers the guarantee to the original costs:
\[
R_{i,c}^k(X_i)
\le
\left(2+\frac1k+\varepsilon\right)c_i(X_j).
\]

It remains to show that
$\log(\widehat C_{\max}/\widehat C_{\min})$
is polynomially bounded in the input size.
Recall that $L_c$ denotes the total binary
encoding length of the input cost matrix.

If all original costs are positive, then
$\widehat C_i(g)=c_i(g)$.
Since every positive rational input cost has
numerator and denominator of at most $L_c$ bits,
we have
\(
2^{-L_c}
\le \widehat C_{\min}
\le \widehat C_{\max}
\le 2^{L_c}.
\)
Consequently,
\[
\log\frac{\widehat C_{\max}}{\widehat C_{\min}}
=O(L_c).
\]

Otherwise, let
\(
C_{\max}^{\mathrm{int}}
=\max_{i\in N,g\in M}C_i(g).
\)
Each row-wise scaling factor $D_i$ is the least
common multiple of the input denominators in
row $i$, and hence has polynomial bit length
in $L_c$. Consequently, every scaled integer
cost $C_i(g)$, and hence
$C_{\max}^{\mathrm{int}}$, has polynomial
bit length in $L_c$.

Recall that
\(
\widehat C_i(g)=C_i(g)+\xi,
\xi=\varepsilon/32m.
\)
Since the scaled costs are nonnegative integers
and at least one original cost is zero, we have
\(
\widehat C_{\min}=\xi,
\widehat C_{\max}=C_{\max}^{\mathrm{int}}+\xi.
\)
Therefore,
$$\log\frac{\widehat C_{\max}}{\widehat C_{\min}}
=
\log\frac{C_{\max}^{\mathrm{int}}+\xi}{\xi}=
\log\left(
1+\frac{32mC_{\max}^{\mathrm{int}}}{\varepsilon}
\right).
$$
Since $C_{\max}^{\mathrm{int}}$ has polynomial
bit length in $L_c$, this logarithmic ratio is
polynomially bounded in $L_c$ and
$\log(1/\varepsilon)$.
Thus, in both cases,
\(
\log(\widehat C_{\max}/\widehat C_{\min})
=
\operatorname{poly}
\left(L_c,\log(1/\varepsilon)\right).
\)

By Theorem~\ref{thm:flow}, the MFCP stage can be
computed exactly in polynomial time.
The sorted-slot rounding, packet decomposition,
and reallocation procedures also run in polynomial
time. The BRBI branch is polynomial as well.
Therefore, Algorithm~\ref{alg:main} runs in
time polynomial in $n$, $m$, $L_c$, and
$1/\varepsilon$.

\end{proof}

\section{BRBI Proof}\label{app:brbi-proof}

This appendix proves the BRBI guarantee from Section~\ref{sec:brbi}. The routine's outline and guarantee are stated in the main text.

\begin{lemma}[Block insertion]\label{lem:brbi-local}

Suppose every current bundle contains at least $k$ chores, and let $B$ be the bundle cheapest to agent $i$. A new block $E$ contains $r$ chores, each no more costly to $i$ than any chore already assigned. Then
\[
R_i^k(B\cup E)\le\max\{1,r/k\}\,c_i(B).
\]
\end{lemma}

\begin{proof}

If $r\le k$, the $k$ cheapest chores can include all of $E$, so the residual cost is at most $c_i(B)$. If $r>k$, let $a=\max_{e\in E}c_i(e)$. Since $|B|\ge k$ and every chore in $B$ costs at least $a$, we have $c_i(B)\ge ka$. After deleting the $k$ cheapest chores, the added residual cost is at most $(r-k)a$, so the total is at most $c_i(B)+(r-k)a\le(r/k)c_i(B)$.
\end{proof}

\begin{proof}[Proof of Theorem~\ref{thm:brbi}]

The claim is immediate when $m\le kn$. Reverse-order reservation ensures that, when agent $i$ is processed in forward order, each chore in $E_i$ costs no more than any chore already assigned. Indeed, the assigned chores belong to base bundles or to blocks $E_s$ for $s<i$, and all were still available when $E_i$ was reserved.

A second invariant is that every unprocessed agent always holds an unexpanded base bundle of size $k$. Any swap affecting her gives her another unexpanded base bundle. Let $b_i$ be the cheapest bundle cost before inserting $E_i$. By Lemma~\ref{lem:brbi-local}, the residual cost of agent $i$ after insertion is at most $\max\{1,r_i/k\}b_i$.

The contents of each physical bundle only increase thereafter, so every final bundle costs agent $i$ at least $b_i$. If agent $i$ later loses her expanded bundle, she receives a base bundle of size $k$; if it is swapped again, she again receives a pure base bundle, leaving zero residual cost. Thus every agent satisfies the claimed bound. Finally, $\max_i r_i=\lceil(m-kn)/n\rceil$. The implementation uses only selection, cost comparisons, and bundle swaps, all in polynomial time.
\end{proof}

\section{Pipeline Tightness}\label{app:tightness-proof}

\begin{theorem}[Tightness of the fixed pipeline]\label{thm:tightness}
For every $k\ge1$ and integer $r\ge3$, set $t=1/[100r(k+1)]$. There exists a positive rational instance $I_{k,r}$ on which MFCP terminates in its first round and every output $Y$ of legal maximum-flow, slot-tie-breaking, and perfect-matching choices satisfies
\[
\rat_k(Y)\ge2+\frac1k-\frac6r.
\]
Thus the output ratio of this specified pipeline approaches $2+1/k$ as $r\to\infty$.
\end{theorem}

This appendix constructs the instance witnessing asymptotic tightness of the specified pipeline and gives a near-exact allocation for the same instance.

Take $n=2r$, $D=100kr$, and $\sigma=1+2t$. Partition the agents into two groups $A,B$, each of size $r$. Define
\[ L=1-\beta+\frac\beta r,\qquad e_A=\frac LD, \]
and
\[ (P,h)=\begin{cases}(1-\rho/r,\rho),&k=1,\\(1-\beta/r,\rho-kP/D),&k\ge2,\end{cases}\qquad e_B=\frac PD. \]
Each chore has a price parameter $p_g$ and a low-cost recipient set $T_g$, and its cost is defined by
\[c_i(g)=\begin{cases}p_g,&i\in T_g,\\\sigma p_g,&i\notin T_g.\end{cases}\]
\begin{table}[t]
\caption{Fixed-pipeline lower-bound instance (each agent has $D$ private chores).}\label{tab:tight-instance}
\centering\normalsize
\begin{tabularx}{\textwidth}{@{}Xccc@{}}
\toprule
Chore type & Count & $p_g$ & $T_g$\\
\midrule
Private chores of group $A$ & $rD$ & $e_A$ & owner\\
Public chores of group $A$ & $r-1$ & $\beta$ & $A$\\
Private chores of group $B$ & $rD$ & $e_B$ & owner\\
Public chore of group $B$ & $1$ & $h$ & $B$\\
\bottomrule
\end{tabularx}
\end{table}

\begin{proposition}\label{prop:good-allocation}
The same instance $I_{k,r}$ admits a complete allocation $Z$ with $\rat_k(Z)\le1+2t$.
\end{proposition}

\begin{proof}[Proof of Theorem~\ref{thm:tightness}]

First, $0<L,P<1$, so $ke_A,ke_B<1/(100r)$. For $k=1$, $h=\rho<\beta$. For $k\ge2$, $t\le1/900$, and
\begin{align*}
\rho-\beta
&=\frac{k-1}{k+1}-t\left(1+\frac4{k+1}\right)\\
&\ge\frac13-\frac7{2700}>\frac1{100r}.
\end{align*}
Therefore $h>\beta$ and $\beta+(k-1)e_A<\rho$. In both cases, the capped price of a public chore is strictly larger than that of a private chore.

Let $q_h=\min\{h,\beta\}$. The construction satisfies
\[
rL+(r-1)\beta=r,\qquad rP+q_h=r.
\]
With initial weights $v_i=1$, the algorithm's base prices equal the parameters $p_g$, and the water level is $1$. Since $\sigma>1+t$, the eligible edges are exactly those with $i\in T_g$. Private chores are assigned entirely to their owners, and public chores are split evenly within their group, giving every agent capped income exactly $1$. Thus the first maximum flow has value $n$, and all eligible edges even satisfy exact MPB.

For any full flow, each private chore must be assigned entirely to its unique eligible recipient. The total public fractional share received by each agent in group $A$ is at most
\[
\frac{r-1}{r}+\frac{t}{n\beta}<1;
\]
Each agent in group $B$ receives public fractional share at most $1/r+t/(nq_h)<1$. Public shares appear first in the sorted order and hence occupy each agent's first slot. Any matching assigns at most one public chore to each agent. Thus some $u\in A$ receives no public chore and $p(Y_u)=L$; the unique public chore in group $B$ is assigned to some $v\in B$, whose bundle price is $p(Y_v)=P+h$.

We next show that the envy-repair stage leaves the allocation unchanged. If $k=1$, all prices are at most $\beta$, so no packets are formed. If $k\ge2$, the total price of the $k$ highest-price chores in each of the four possible bundle types is at most
\[
ke_A,\quad \beta+(k-1)e_A,\quad ke_B,\quad
h+(k-1)e_B=\rho-e_B.
\]
All these values are strictly below $\rho$, and the $(k+1)$-st chore is also priced below $\rho$. Hence every agent is in the zero-packet case.

Since $h>e_B$, the comparison ratio from $v$ to $u$ is
\begin{equation}\label{eq:forced-ratio}
F_k(r)=\frac{R_v^k(Y_v)}{c_v(Y_u)}
=\frac{P+h-ke_B}{\sigma L}.
\end{equation}
Let $d=k/(k+1)$. As $r\to\infty$, $L\to d$, $P\to1$, $h\to d$, $ke_B\to0$, and $\sigma\to1$, so $F_k(r)\to(1+d)/d=2+1/k$.

For a finite-parameter bound, use $\rho\ge d-2t$, $P\ge1-1/r$, $L\le d+1/(2r)$, $2ke_B\le1/(50r)$, and $2t\le1/(100r)$ to obtain
\[
P+h-ke_B\ge1+d-\frac{103}{100r},\qquad
\sigma L\le d+\frac{51}{100r}.
\]
Let $Q=(1+d)/d\le3$, then
\begin{align*}
Q-\frac{1+d-103/(100r)}{d+51/(100r)}
&=\frac{103+51Q}{100r\bigl(d+51/(100r)\bigr)}\\
&\le\frac{512}{100r}<\frac6r.
\end{align*}
Therefore, every legal output satisfies the stated lower bound.
\end{proof}

\begin{proof}[Proof of Proposition~\ref{prop:good-allocation}]

Ignore the support restriction and sort all chores in nonincreasing order of $p_g$, assigning each in turn to a currently cheapest bundle. For each nonempty final bundle $Z_i$, let $g$ be the last chore assigned to it. Since chores are processed in nonincreasing order of price, $g$ is a cheapest chore in $Z_i$. When $g$ is assigned, the price of $Z_i\setminus\{g\}$ is minimum among all current bundles. Since the prices of other bundles can only increase afterward, we have $p(Z_i\setminus\{g\})\le p(Z_j)$ for every $j$. Hence the allocation is price-EFX: $R_p^1(Z_i)\le p(Z_j)$. Since $p_g\le c_i(g)\le\sigma p_g$,
\[
R_i^k(Z_i)\le R_i^1(Z_i)\le\sigma R_p^1(Z_i)
\le\sigma p(Z_j)\le\sigma c_i(Z_j).
\]\end{proof}
\section{Ten-Chore Lower Bound}\label{app:lower-proof}
\begin{theorem}[EFX existence lower bound]\label{thm:lower}
There is a nonnegative additive-cost instance with four agents and ten chores whose optimal EFX approximation factor is \(\gamma>11/10\).
\end{theorem}

\subsection{Construction and Parameters}
Let $\gamma$ denote the real root greater than $1$ of
$
2\gamma^4+2\gamma^3-2\gamma^2-2\gamma-1=0.
$ The polynomial is strictly increasing on $[1,\infty)$ and
has opposite signs at $11/10$ and $111/100$.
Thus, $\gamma$ is uniquely determined and satisfies
$11/10<\gamma<111/100$.
Set $
p=\gamma-1, \mu=\frac1\gamma-2p,
\nu=\gamma(1+2p)-\mu.
$
The defining equation gives
\begin{equation}\label{eq:lower-identities}
1+p=\frac1{\mu+2p}=\frac{\mu+\nu}{1+2p}
=\frac{\mu+p}{\nu+p}=\gamma.
\end{equation}
The parameter definitions and the above interval also imply
\begin{equation}\label{eq:lower-aux}
\begin{gathered}
0<p<\nu<\mu<1,\quad
\mu+p<1,\quad 2p+\nu<1,\quad \nu>2\gamma p,\\
3\mu>\gamma,\quad 3\mu+\nu>2\gamma,\quad
\gamma(3\mu+\nu+2p)<4.
\end{gathered}
\end{equation}
The instance has four agents and ten chores:
\[
A=\{A_1,A_2,A_3\},\quad B=\{B_0,B_1,B_2\},\quad
C=\{C_0,C_1,C_2\},\quad \{D\}.
\]
Their per-chore costs are shown in Table~\ref{tab:exist-instance}.
\begin{table}[t]
\caption{Per-chore costs in the four-agent, ten-chore EFX lower-bound instance.}\label{tab:exist-instance}
\centering
\begin{tabular}{@{}lcccccc@{}}
\toprule
agent & each \(A\) chore & \(B_0\) & \(B_1,B_2\) & \(C_0\) & \(C_1,C_2\) & \(D\)\\
\midrule
\(1,2\) & \(1\) & \(0\) & \(p\) & \(\mu\) & \(\mu\) & \(\nu\)\\
\(3,4\) & \(1\) & \(\mu\) & \(\mu\) & \(0\) & \(p\) & \(\nu\)\\
\bottomrule
\end{tabular}
\end{table}

\paragraph{Positive integer-cost check.}
Replacing \(1,0,p,\mu,\nu\) by \(10000001,1,1025785,7018082,6269720\), respectively, gives a strictly positive integer instance. Exact enumeration of the \(128000\) type-count allocations, covering all \(4^{10}\) labelled allocations, gives optimal factor
\[
\frac{10000001}{9069653}>\frac{11}{10}.
\]
This finite check is supplementary; the algebraic lower bound is proved below.

\subsection{Excluding All Complete Allocations Below
\texorpdfstring{$\gamma$}{gamma}}
\begin{proof}[Proof of Theorem~\ref{thm:lower}]
Suppose, for a contradiction, that a complete \(\alpha\)-EFX allocation exists for some \(1\le\alpha<\gamma\). Call \(B\cup C\cup\{D\}\) the ordinary chores. For either agent type, their costs form the multiset
\[
\{0,p,p,\nu,\mu,\mu,\mu\},
\]
with total \(3\mu+\nu+2p\).

\paragraph{Bundles containing multiple \(A\)-chores.}
If a bundle contains three \(A\)-chores, or two \(A\)-chores and an ordinary chore, its residual cost is at least \(2\). At least two other bundles are then ordinary-only, and each must cost that same agent strictly more than \(2/\gamma\). Thus the total cost of all ordinary chores to that agent exceeds \(4/\gamma\), contradicting \(\gamma(3\mu+\nu+2p)<4\). Thus the only possible bundle containing multiple \(A\)-chores is the pure bundle \(\{A,A\}\).

By symmetry, suppose agent \(1\) holds \(\{A,A\}\). Her residual cost is \(1\), so at least two ordinary-only bundles must each cost agent \(1\) more than \(1/\gamma=\mu+2p\). An ordinary-only bundle containing at most one chore from \(C\cup\{D\}\) costs agent \(1\) at most \(\mu+2p\), even if it receives all \(B\)-chores. Hence each of the two bundles contains at least two chores from \(C\cup\{D\}\). Since there are exactly four such chores, all of them go to these two bundles, and the single-\(A\) bundle can contain only \(B\)-chores.

\paragraph{The single-\(A\) bundle is held by the other type.}
One ordinary-only bundle is held by agent \(2\). If it also contains a \(B\)-chore, her residual cost is at least \(\mu+\nu\), while the single-\(A\) bundle costs her at most \(1+2p\). By \eqref{eq:lower-identities}, this gives a ratio of at least \(\gamma\), a contradiction. Thus agent \(2\) receives no \(B\)-chore. From the other type's perspective, agent \(2\)'s bundle costs at most \(\nu+p\). If the single-\(A\) holder receives any \(B\)-chore, her residual cost is at least \(1\), and
\[
1>\mu+p=\gamma(\nu+p),
\]
again contradicting \(\alpha<\gamma\). Therefore the single-\(A\) bundle is pure. All three \(B\)-chores must then go to the other ordinary-only bundle, which also contains a cheap \(C\)-chore. After deleting its cheapest chore, all three \(B\)-chores remain, giving residual cost at least \(3\mu>\gamma\), a contradiction with the pure single-\(A\) bundle.

\paragraph{The single-\(A\) bundle is held by the same type.}
Both ordinary-only bundles are now held by agents of the other type. Let \(P\) contain two \(C\)-chores and \(Q\) contain one \(C\)-chore and \(D\). If \(P\) has no \(B\)-chore, its cost to its holder is at most \(2p\), whereas the residual cost of \(Q\) is at least \(\nu>2\gamma p\), a contradiction. Thus \(P\) contains a \(B\)-chore. If \(Q\) has no \(B\)-chore, its cost is at most \(\nu+p\), whereas the residual cost of \(P\) is at least \(\mu+p=\gamma(\nu+p)\), again a contradiction. Hence both \(P\) and \(Q\) contain a \(B\)-chore.

If the single-\(A\) bundle contains no \(B\)-chore, it is a pure singleton and \(P,Q\) contain all ordinary chores. Both bundles contain a cheap \(C\)-chore, so their residual costs sum to at least \(3\mu+\nu>2\gamma\). But comparison with the pure singleton requires this sum to be at most \(2\alpha<2\gamma\), a contradiction. If the single-\(A\) bundle contains one \(B\)-chore, then \(P,Q\) each contain exactly one. The residual cost of \(Q\) is at least \(\mu+\nu\), while \(P\) costs at most \(\mu+2p<1+2p\); hence the ratio exceeds \((\mu+\nu)/(1+2p)=\gamma\), again a contradiction.

Thus the three \(A\)-chores must be assigned to three distinct agents. Let \(S\) be the unique bundle without an \(A\)-chore.

\paragraph{No \(A\)-holder can value \(S\) too cheaply.}
Suppose an agent \(i\) holding an \(A\)-chore has \(c_i(S)\le\mu+2p=1/\gamma\). If she also holds an ordinary chore, her residual cost is at least \(1\), contradicting comparison with \(S\). Hence she must hold the pure single-\(A\) bundle. Each of the other two $A$-holders can then receive at most
one ordinary chore. Indeed, if either receives at least two ordinary chores, deleting her cheapest chore leaves her $A$-chore and at least one ordinary chore of cost at least $p$, since at most one ordinary chore has zero cost to her. Her residual cost is therefore at least $1+p=\gamma$,
again contradicting comparison with the pure singleton. Consequently, at least five of the seven ordinary chores must lie in \(S\). Among any five or more elements of \(\{0,p,p,\nu,\mu,\mu,\mu\}\), the residual cost after deleting the cheapest is at least
\[
\mu+\nu+2p=\gamma(1+2p)+2p>\gamma,
\]
a contradiction. Therefore every \(A\)-holder values \(S\) at strictly more than \(\mu+2p\). Since the three \(A\)-holders include both types, both types value \(S\) above this threshold.

\paragraph{The bundle without an \(A\)-chore gives the final contradiction.}
The threshold forces \(S\) to contain at least two chores from \(C\cup\{D\}\) and at least two from \(B\cup\{D\}\). If \(D\in S\), then \(S\) contains a \(B\)-chore and a \(C\)-chore. After deleting the cheapest chore, the \(C\)-chore of cost \(\mu\) and \(D\) of cost \(\nu\) remain, so \(R_1(S)\ge\mu+\nu\). At most two \(C\)-chores remain outside \(S\), so one of the three \(A\)-holders has no \(C\)-chore and no \(D\). Her bundle costs agent \(1\) at most \(1+2p\), giving a ratio of at least \(\gamma\).

If \(D\notin S\), then \(S\) contains at least two \(B\)-chores and two \(C\)-chores, so \(R_1(S)\ge2\mu+p\). At most one \(C\)-chore and one \(D\) remain outside \(S\), so one \(A\)-bundle contains neither. At most one \(B\)-chore remains outside \(S\), and this comparison bundle costs at most \(1+p\). Since \(2\mu+p>\mu+\nu\) and \(1+p<1+2p\), the ratio is strictly greater than
\[
\frac{\mu+\nu}{1+2p}=\gamma.
\]
Both cases contradict \(\alpha<\gamma\). Thus every complete allocation has EFX approximation factor at least \(\gamma\).

\end{proof}

\subsection{An Allocation Attaining the Lower Bound}
Take
\[
Z_1=B\cup\{D\},\quad Z_2=\{A_1\},\quad
Z_3=\{A_2,C_1\},\quad Z_4=\{A_3,C_0,C_2\}.
\]
Agent \(1\)'s residual cost is \(2p+\nu<1\), and her cheapest comparison bundle costs \(1\). Agent \(2\)'s residual cost is zero. Agent \(3\)'s residual cost is \(1\), and her cheapest comparison bundle costs \(1\). Agent \(4\)'s residual cost is \(1+p=\gamma\), and her cheapest comparison bundle costs \(1\). Hence \(\rat(Z)=\gamma\). Hence, the optimal EFX approximation factor of
$I_\gamma$ is $\gamma>11/10$.

\end{document}